\documentclass[article,twocolumn]{revtex4}
\usepackage{graphicx}
\usepackage{color}

\usepackage{dcolumn}
\usepackage{bm}
\usepackage{amssymb,amsmath}
\usepackage{color}

\definecolor{DarkGreen}{rgb}{0,0.6,0.2}

\newcommand{\ket}[1]{\left| #1 \right\rangle}
\newcommand{\bra}[1]{\left\langle #1 \right|}

\newcommand{\Tr}{\mathrm{Tr}}

\newcommand{\hf}{\frac{1}{2}}

\begin{document}
\title{Detecting drift of quantum sources: not the de Finetti theorem}
\author{Lucia Schwarz} 
\author{S.J. van Enk}
\affiliation{Department of Physics\\
Oregon Center for Optics\\University of Oregon, Eugene OR 97403}
\begin{abstract}
We propose and analyze a method to detect and characterize the drift of a nonstationary quantum source. It generalizes a standard measurement for detecting phase diffusion of laser fields to  quantum systems of arbitrary Hilbert space dimension, qubits in particular.
We distinguish diffusive and systematic drifts, and examine how quickly one can determine that a source is drifting. We show that for single-photon wavepackets our measurement is implemented by the Hong-Ou-Mandel effect.
\end{abstract}

\maketitle

Ever since its first experimental implementation in 1993 \cite{Raymer1993}, 
quantum tomography (or, more generally, quantum-state estimation \cite{Blume2010}) has become an important tool in the field of quantum information science (for reviews, see \cite{Paris2004,Lvovsky2009}).
From the results of different measurements on many instances of
a quantum system, one infers a density operator, $\rho_0$ (plus ``error bars'') that describes best (in some more or less well-defined sense) the state of each instance. 
One may well wonder why one assigns just {\em one} density operator. A crucial role in this context is played by the de Finetti theorem (for details, on both the infinite and finite versions of the theorem, see \cite{Caves2002, KoeRen2005}): if
one has an extendible permutation-invariant sequence of $N$ quantum systems, then one can assign a quantum state of the form
\begin{equation}
\rho^{(N)}=\int d\rho P(\rho)\rho^{\otimes N}
\end{equation}
to the collection of $N$ systems, with
$P(\rho)$ a probability density over density operators. Quantum tomography then succeeds in making the distribution $P(\rho)$ more and more narrow, sharply peaked around some $\rho_0$. In fact, in the limit of $N\rightarrow\infty$, one has $P(\rho)\rightarrow \delta(\rho-\rho_0)$.

We are interested here in the case where the assumption of permutation invariance does not hold, and where, consequently, the de Finetti theorem does not apply. The most relevant case is that of a (slowly) drifting source.
For example, it is well known that a laser displays phase diffusion: when one considers two light pulses emitted by the same laser with a short time delay $\tau$ between them, there will be a (random) phase difference whose average magnitude increases with $\tau$.  Of course, even in this case, one could average over all emitted light pulses, say $N$ instances, to arrive at a single average density matrix.
Indeed, if done correctly the averaging procedure restores the permutation invariance, but (i) the average density matrix depends on the number $N$, and (ii) the averaging procedure throws away potentially useful information. For example, if we are interested in the purity of our quantum states, the single state estimate will be too conservative.
 
In this paper we set ourselves the task of figuring out how one could detect 
whether (and how) a source is drifting. 
In principle, for detecting drift one could still use a variant of quantum tomography: for example, we split our $N$ quantum systems into two groups of size $N/2$ each: the first half (chronologically) and the second half. For each we estimate a single density matrix: and if the difference between the two estimates is (not) statistically significant then we conclude our source is (not) likely drifting. This method works to some extent, but is still subject to the same two objections mentioned above. Moreover, it has been known for a few decades that for the detection of given physical quantities (such as a particular matrix element of the density matrix) a targeted method is always superior to performing full tomography \cite{Dariano95}.
Therefore, we propose and analyze a different method directly targeted at detecting drift. A difference with the above-mentioned method \cite{Dariano95} is that we consider a quantity determined by {\em pairs} of density matrices.

Consider what one would measure to detect phase diffusion of a (pulsed) laser in the special (but relevant) case where one assumes the laser pulses can be described by coherent states with some fixed (and known) amplitude but a diffusing phase (relative to some phase standard). One would take pairs of the output laser pulses, and split them on a 50/50 beamsplitter in such a way that one particular output would be the vacuum if their phase difference, $\delta\phi$, would be zero. That output's intensity is then $I=|\alpha|^2|1-\exp(i\delta\phi)|^2/2$, if $|\alpha|$ is the amplitude of the laser pulses. Thus measuring this intensity determines the phase difference directly. 

Now how do we generalize this measurement to arbitrary quantum systems (in particular, qubits)? We first note
that the intensity $I$ can also be written in terms of the {\em overlap} between the two input states, call them $\rho=|\alpha\rangle\langle\alpha|$ and $\rho'=|\alpha'\rangle\langle\alpha'|$, since $\exp(-2I)=\Tr(\rho\rho')$ in this case.
Thus, our choice of generalization will be to measure the overlap between pairs of instances of quantum systems from one and the same source.
In other words, we propose to measure the 
swap operator $\hat{V}$, defined in terms of basis vectors $\{|i\rangle\}$ and $\{|j\rangle\}$ of the two (isomorphic) Hilbert spaces of two instances numbered $m$ and $n$ from our source by
\begin{equation}
\hat{V}=\sum_i\sum_j |i\rangle_m\langle j|\otimes|j\rangle_n\langle i|.
\end{equation}
 The expectation value of $\hat{V}$ equals the overlap
\begin{equation}
\Tr (\rho_m\otimes \rho_n \hat{V})=\Tr (\rho_m\rho_n).
\end{equation}
(Note the left-hand side contains the tensor product of two density operators, the right hand-side their matrix product.)
Each measurement of $\hat{V}$ yields one of its two eigenvalues, $\pm 1$, and so only after multiple measurements will one obtain a statistical  estimate of the overlap.
[And, as a bonus, {\em if} the two density matrices are identical, then this measurement in fact measures the purity \cite{Filip02,Ekert02}, $P=\Tr(\rho_m^2)$.] 
By comparing the overlap between adjacent copies, where $|m-n|=1$ with
the overlap between outputs that are farther apart, $|m-n|>1$, we obtain information about whether the source is drifting: if the source is not drifting, the overlap is independent of $|m-n|$. 

In order to infer more detailed information about the character of the drift or diffusion (beyond the mere statement that the source is or is not stationary), we need some simplifying assumptions about the sequence of states (the space of {\em all} possible output states of $N$ copies is too large to be either measurable or tractable).
Here we make the following two  assumptions (one of which has been implicitly used already in the above description): (a) the states are independent, (b) the drifting process is Markovian, such that the overlap between two copies $m$ and $n$ only depends on $|m-n|$. 
So we write the state of $N$ systems produced by our quantum source as a tensor product \footnote{It is an interesting question whether this form can be {\em derived} from assumptions about {\em approximate} permutation invariance (plus extendability) of the sequence (private communication with C.~Caves, many years ago). Also note that for laser pulses emitted by a phase-diffusing laser their state can indeed be written in this form, modulo two subtleties discussed in \cite{VanEnkFuchs2002}.}
\begin{equation}
\rho^{(N)}=\rho_N\otimes\rho_{N-1}\ldots\otimes\rho_1.
\end{equation}
In this case, we get
\begin{equation}
V_{nm}\equiv{\rm Tr}[\rho_n\otimes\rho_m\hat{V}]
=\frac{1}{2}[P_n+P_m]-\frac{1}{2}{\rm Tr}[\Delta_{mn}^2],
\end{equation}
where $P_k={\rm Tr}[(\rho_k)^2]$ is the purity of system $k$, and $\Delta_{mn}$ is the difference between the two  states $m$ and $n$:
\begin{equation}
\Delta_{mn}=\rho_m-\rho_n.
\end{equation}
As special cases of nonstationary sources we consider both diffusion and systematic drift. Consider a (Markovian) process where
\begin{equation}\label{process}
\rho_{n+1}= U_r\rho_n U_r^\dagger.
\end{equation}
Diffusive drift occurs when $U_r$ is a {\em random} unitary matrix, picked from some distribution; a systematic drift occurs when $U_r$ is fixed. In either case, the purity of  $\rho_n$ is independent of $n$:
$
P_n=P_m\equiv P_1.
$
We can take stochastic averages over the random distribution of unitaries, which we will indicate by a bar, to get
\begin{equation}
\overline{V}_{nm}=P_1-\frac{1}{2}{\rm Tr}
\overline{[\Delta_{nm}^2]}.
\end{equation}
In case the drift process is purely a systematic drift, each unitary $U_r$ is the same, and we get
\begin{equation}\label{D1}
\Tr \overline{[\Delta_{nm}^2]}=|n-m|^2 D_1
\end{equation} for some {\em drift constant} $D_1$.  

\begin{figure}[h]
	\includegraphics[width=.45\textwidth]{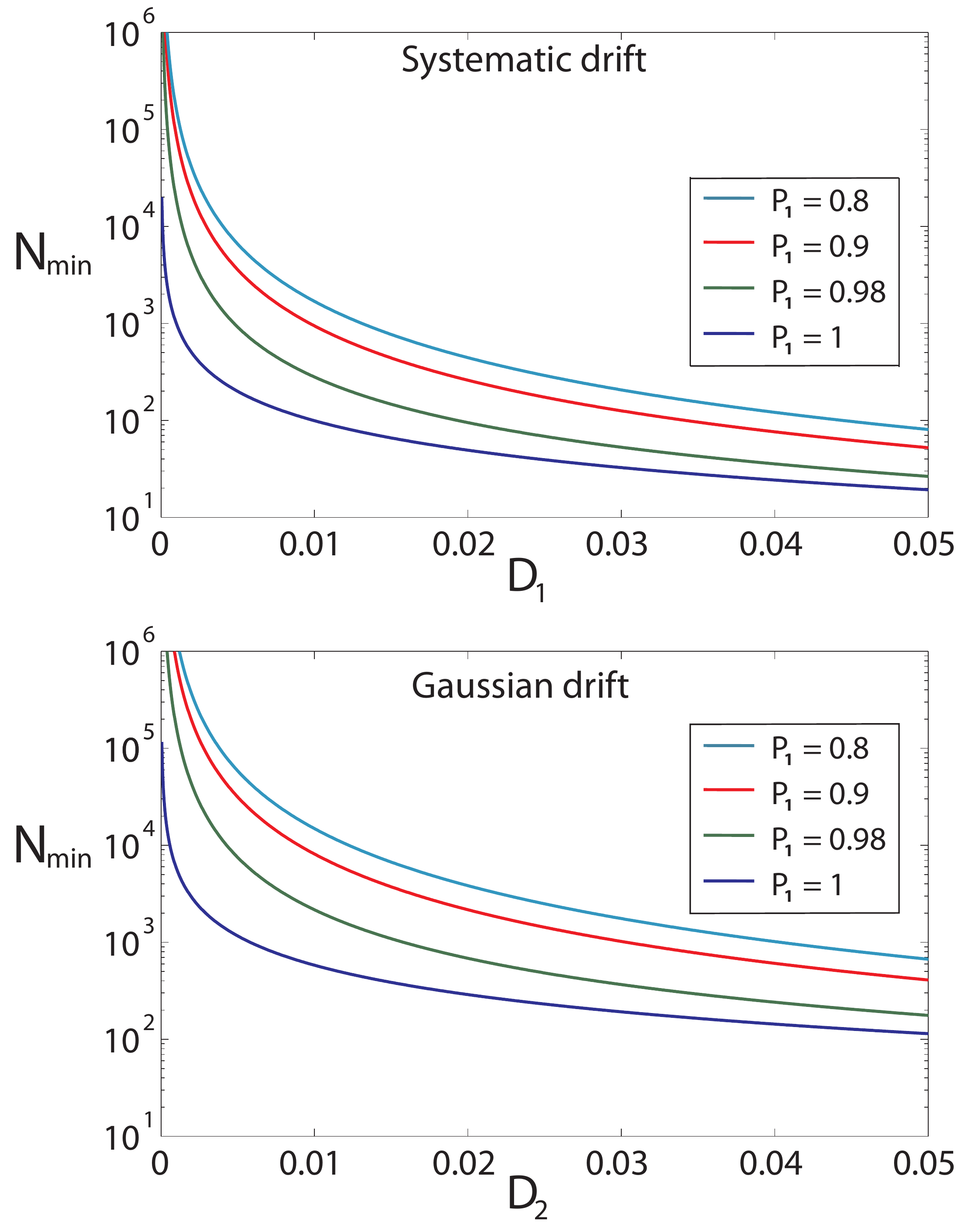}
		\caption{How many measurements do we need to figure out that a source is drifting? Obviously, the larger the drift is (as measured by the parameter $D_2$ for diffusive drift or $D_1$ for systematic drift), the fewer measurements we need. Top: systematic drift, bottom: diffusive drift. }
	\label{howmanyn}
\end{figure}

If the process that changes the states $\rho_n$ is diffusive (for example, the random distribution of $U_r$ is a Gaussian centered around the identity), then we get a linear relationship between the overlap and the distance $|n-m|$,
\begin{equation}\label{D2}
\Tr \overline{[\Delta_{nm}^2]}=|n-m| D_2,
\end{equation}
for some {\em diffusion constant} $D_2$ \footnote{Equations (\ref{D1}) and (\ref{D2}) are informative only for states sufficiently far away from the completely mixed state, since the latter state is invariant under (\ref{process}), and so $D_1$ and $D_2$ would have to be zero; such states sufficiently far from the mixed state are, of course, also the most relevant for quantum information processing purposes, and we thus assume this condition to be fulfilled.}.
In this case, measuring the swap operator between neighboring copies, for which $|n-m|=1$ and on copies with $|n-m|=2$ gives us both the purity $P_1$ and the diffusion constant $D_2$:
\begin{eqnarray}
P_1 &=& 2 \Tr \overline{[\Delta_{n,n+1}^2]} - \Tr \overline{[\Delta_{n,n+2}^2]},\\
D_2 &=& \Tr \overline{[\Delta_{n,n+2}^2]} - \Tr \overline{[\Delta_{n,n+1}^2]}.
\end{eqnarray}
(And similar relations hold when the drift is purely systematic.)
One way to check which sort of drift process one actually has, diffusive, systematic, or a combination thereof, is to measure in addition the quantity $\Tr \overline{[\Delta_{n,n+3}^2]}$, and
then calculate the ratio
\begin{equation}
\alpha\equiv\frac{\Tr \overline{[\Delta_{n,n+2}^2]} - \Tr \overline{[\Delta_{n,n+1}^2]}}{\Tr \overline{[\Delta_{n,n+3}^2]}-\Tr \overline{[\Delta_{n,n+2}^2]}}.
\end{equation}
If the ratio is 1, one has  a purely diffusive process, if $\alpha=3/5$ one has a systematic drift, and in all cases in between one has both diffusive and systematic drifts.
To see how the number $\alpha$ is determined when there is a combination of systematic and diffusive drifts, let us consider the simplest case of a qubit source.
We model the drift process with a unitary matrix $U_r = \exp(i\delta\vec{r}\cdot\vec{\sigma})$, with $\delta\ll 1$, $\vec{\sigma}$ a vector containing the three Pauli matrices, and a random vector $\vec{r}$ that consists of both a diffusive part and a systematic part, 
\begin{equation}
\vec{r} = p\, \vec{r}_{\rm const} + (1-p)\, \vec{r}_{\rm diffusive}
\end{equation}
with a normally distributed random vector $\vec{r}_{\rm diffusive}$ and a constant (unit) vector $\vec{r}_{\rm const}$ (and $0\leq p\leq 1$). In this case $\alpha$  depends on $p$ and on the ratio of the two constants $D_1$ and $D_2$, with  $D_1 = \delta^2\Tr [\vec{r}_{\rm const}\cdot\vec{\sigma}, \overline{\rho}]^2$ and $D_2=\delta^2\Tr \overline{[\vec{r}_{\rm diffusive}\cdot\vec{\sigma}, \rho]^2}$:
\begin{equation}
\alpha = \frac{-p^2+\frac{D_1}{D_2} (-3p^2+6p-3)}{-p^2+\frac{D_1}{D_2}(-5p^2+10p-5)}.
\end{equation}
The next question  we consider is, given a source of quantum states, how quickly can we determine (by measuring the swap operator) whether the source is drifting? Let us
first consider the case of pure diffusive drift. We could, for example, measure the swap operator between states that are 1 step and 2 steps apart, respectively, and see if the two numbers are equal or not.  Suppose we write the overlaps as 
\begin{align} \label{assumption}
& \overline{{\rm Tr} (\rho_n \rho_{n+1})} = P_1-D_2/2, \notag \\
& \overline{{\rm Tr} (\rho_n \rho_{n+2})} = P_1-D_2. \\
& \ldots \notag
\end{align}
Suppose we have $N$ data sets of the measurements of both ${\rm Tr} (\rho_n \rho_{n+1})$ and ${\rm Tr} (\rho_n \rho_{n+2})$. We get the measured frequencies $f^{\pm}_1$ and $f^{\pm}_2$, respectively, of the measurement outcomes $\pm 1$ in the two cases. The average values are 
\begin{eqnarray} \label{purities}
V_1 := \overline{{\rm Tr} (\rho_n \rho_{n+1})} = f_1^+ - f_1^-, \notag \\
V_2 := \overline{{\rm Tr} (\rho_n \rho_{n+2})} = f_2^+ - f_2^-,
\end{eqnarray}
and the standard error bars  are (for large enough $N$)
\begin{eqnarray}
\Delta V_1 = 2 \sqrt{\frac{f_1^+ f_1^-}{N}}, \notag \\
\Delta V_2 = 2 \sqrt{\frac{f_2^+ f_2^-}{N}}.
\end{eqnarray}
To decide that the source is drifting,   the values of $V_1$ and $V_2$ should \emph{not} overlap within their error bars. The necessary condition for that is
\begin{equation} \label{driftcondition}
\frac{1}{2} \Delta V_1 + \frac{1}{2} \Delta V_2 < V_1 - V_2 = D_2/2.
\end{equation}
From assumption (\ref{assumption}) and Eqns.~(\ref{purities}), we can write the frequencies in terms of $D_1$ and $P_1$ 
\begin{align}
f_1^+ = \hf (1 + P_1 - \frac{D_2}{2}); \; & f_1^- = \hf (1 - P_1 + \frac{D_2}{2}), \notag \\
f_2^+ = \hf (1 + P_1 - D_2); \;           & f_2^- = \hf (1 - P_1 + D_2),
\end{align}
and using all this in Eq.~(\ref{driftcondition}) we can solve for the minimum necessary number of measurements:
\begin{align}
N_{\min} = [ \frac{1}{D_2} (& \sqrt{(1+P_1-\frac{D_2}{2})(1-P_1+\frac{D_2}{2})} \notag \\ +& \sqrt{(1+P_1-D_2)(1-P_1+D_2)} ) ]^2.
\end{align}
To detect a systematic drift, a similar calculation gives 
\begin{align}
N_{\min} = [ \frac{1}{3 D_1} (& \sqrt{(1+P_1-\frac{D_1}{2})(1-P_1+\frac{D_1}{2})} \notag \\ +& \sqrt{(1+P_1-2 D_1)(1-P_1+2 D_1)} ) ]^2.
\end{align}
The number of measurements needed, for both diffusive and systematic drifts, is depicted in Fig. \ref{howmanyn} for various values of $P_1$.

\begin{figure}[h]
	\includegraphics[width=.5\textwidth]{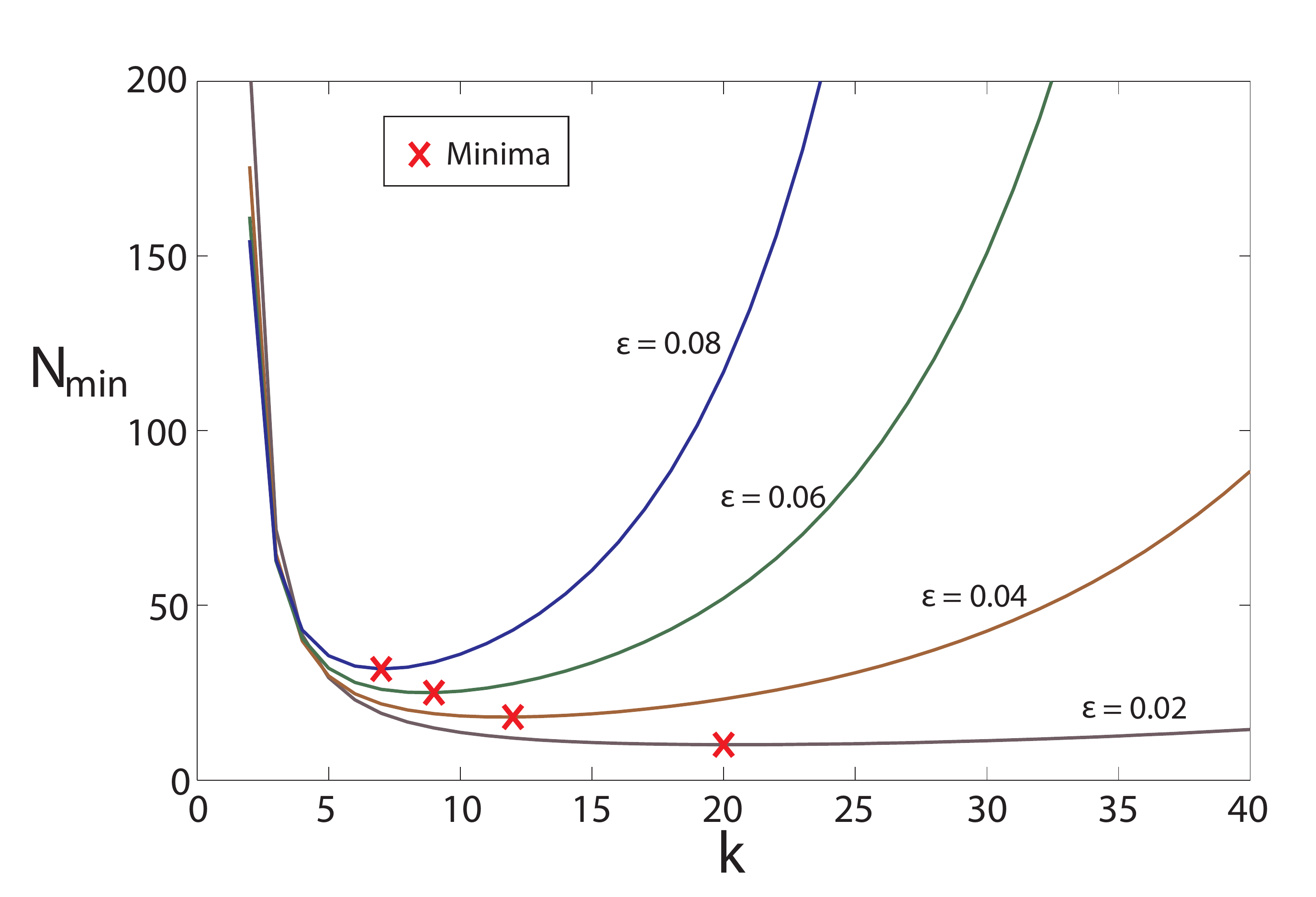}
	\caption{This plot shows the minimal number of measurements needed to detect drift, as a function of the (temporal) distance between states measured, $k$, for various values of the decoherence parameter $\epsilon$, for $P_1 = 1$, $D= 2$ and $D_2 = 0.01$ (see main text for definitions). The minimum of each curve determines the optimal distance $k$ between states to be measured (in addition to distance-1 overlaps). }
	\label{nvsk}
\end{figure}

In principle, one could detect a drifting source a lot faster if one measured the swap operator on states that are $k>2$ steps apart (in addition to measuring states 1 step apart), simply because $|V_k-V_1|$ will be  larger. In an actual experiment, however, the larger the distance between two copies, the longer the earlier copy would have to be stored in memory. We could model the decoherence that the earlier copy undergoes as follows: assume that there is a typical decoherence time scale $\tau$, which, e.g., drives any state towards the totally mixed state. That is, if we keep a system for time $t$, then
$\rho\rightarrow (\exp(-t/\tau)\rho+
(1-\exp(-t/\tau)\openone/D$, with $D$ the dimension of the Hilbert space of our quantum system. Then assume that the time needed to produce one copy is
$\epsilon \tau$
with some (hopefully small) number $\epsilon$. Then we can write the  overlap between states $n$ and $n+k$ as
\begin{align} P_k & = {\rm Tr}(\rho_n \tilde{\rho}_{n+k}) \notag \\
   & = e^{-k\epsilon}{\rm Tr}(\rho_n \rho_{n+k}) + \frac{1}{D}(1-e^{-k\epsilon}){\rm Tr}(\openone \rho_{n+k}).
  \end{align}
The inferred overlap between copies $n$ and $n+k$ follows from the measured $P_k$ by multiplying it with $\exp(k\epsilon)$ (and subtracting a known quantity): so the error bar in the overlap multiplies by the same number. This error thus becomes substantial once $k\epsilon$ becomes of order unity, so that is where we would expect the method to use copies a distance $k$ apart to break down. In Fig.\ref{nvsk} we plot the number of measurements needed for various values of $\epsilon$ as a function of $k$, and we can indeed see that for too large values of $k$,  the required number of measurements increases exponentially with $k$. The optimal $k$, of course, depends on the specific decoherence process, and also on how fast the source is drifting, but seems to be around $k\epsilon\approx 4$ in our example.

Finally, we wish to note that
in the case of two independent single photons, when they are viewed as quantum systems with an infinite-dimensional Hilbert space describing polarization, spectral, and transverse spatial degrees of freedom, the swap operator can in fact be measured via the Hong-Ou-Mandel interference effect \cite{HOM1987}. This can be shown as follows. We consider two single photon wavepackets impinging on two different input ports (denoted A and B) of a 50/50 beamsplitter. We write the two (mixed) input states in terms of creation and annihilation operators $a^{\dagger}$ and $a$ (for port A) and $b^{\dagger}$ and $b$ (for port B) as:
\begin{eqnarray}
& \rho_A = \sum_{kl} p_{kl} a_k^{\dagger} \ket{0} \bra{0} a_l \\
& \rho_B = \sum_{nm} q_{nm} b_n^{\dagger} \ket{0} \bra{0} b_m,
\end{eqnarray}
where the subscripts stand for the mode properties (polarization, frequency etc.) other than their propagation direction. The combined input state is then
$\rho_{\rm in} = \rho_A \otimes \rho_B $. 
This state gets transformed by the 50/50 beamsplitter in the following way: 
\begin{eqnarray}
& \rho_{\rm out} = \sum_{klnm} \frac{p_{kl} q_{nm}}{4} (c_k^{\dagger} + i d_k^{\dagger})(i c_n^{\dagger} + d_n^{\dagger}) \\
& \ket{0} \bra{0} (c_l-id_l)(-ic_m+d_m), \notag
\end{eqnarray}
where $c$ and $d$ now denote operators of the two output ports C and D.
To get the probability $P_{cc}$ of getting a coincidence count, i.e., photo detections at both output ports C and D, we take a partial trace:
\begin{eqnarray}
P_{cc} &  = \sum_{rs} \bra{1_r}_c \bra{1_s}_d \rho_{\rm out} \ket{1_r}_c \ket{1_s}_d \\
    & = \sum_{rs} \bra{0} c_r d_s \rho_{\rm out} c_r^{\dagger} d_s^{\dagger} \ket{0} \notag \end{eqnarray}
This simplifies to 
\begin{equation} P_{cc} = \frac{1}{2} \sum_k p_{kk} \sum_n q_{nn} - \frac{1}{2} \sum_{kl} p_{kl} q_{lk}. \end{equation}
The first two sums are the traces of the density matrices and therefore equal 1. It is easy to see that 
\begin{equation} \mathrm{Tr} (\rho_A \rho_B) = \sum_{klnm} p_{kl} q_{nm} \delta_{nl} \delta_{mk} = \sum_{kl} p_{kl} q_{lk}, \end{equation} 
so that we get the simple relation
\begin{equation} 2 P_{cc} = 1 - \rm Tr(\rho_A \rho_B).\end{equation}
Thus, as announced, the HOM effect measures the overlap between two input states, and hence the swap operator. (And so, if the two single-photon input states are identical, then the HOM interference measurement measures the purity of the input states. 
Note that this is different from the  measurement of single-photon (spectral) purity implemented recently in Ref.~\cite{cassemiro2010}, which also makes use of the HOM effect, but with a known coherent-state input in the other input port.) Of course, the HOM effect has been measured many times in the context of characterizing single-photon sources  (see, e.g., \cite{Santori02,Legero06}), but never, as far as we know, systematically on copies more than the minimum distance apart. We also note that for the polarization degree of single photons, the overlap has been measured \cite{Hendrych03}, following ideas from \cite{Filip02}.

In conclusion, we proposed the measurement of the swap operator as a means to detect the drifting of a quantum source. This measurement complements quantum tomography, which produces an estimate of a {\em single} average density matrix, by partially characterizing how this estimate would change over time, for instance, distinguishing between diffusive and systematic drifts. We also analyzed how many measurements are needed to determine that a source is drifting, including the influence of decoherence on the precise measurement strategy.
We showed the swap measurement  on pairs of single-photon wavepackets is implemented simply by the Hong-Ou-Mandel effect.

\bibliography{Short}

\end{document}